\documentclass{amsart}

\usepackage[centertags]{amsmath}
\usepackage{times}
\usepackage{amsfonts,amssymb}
\usepackage{fullpage}

\newcommand{\ZZ}{\mathbb{Z}}

\newcommand{\NN}{\mathbb{N}}

\newcommand{\chr}{\varphi}

\renewcommand{\u}{\mathcal{U}}

\title{Anomalous Dissipation in a Stochastically Forced Infinite-Dimensional 
  System of Coupled Oscillators} 
\author{Jonathan C. Mattingly, Toufic M. Suidan, Eric Vanden-Eijnden}

\begin{document}

\begin{abstract}
  We study a system of stochastically forced infinite-dimensional
  coupled harmonic oscillators. Although this system formally
  conserves energy and is not explicitly dissipative, we show that it
  has a nontrivial invariant probability measure. This phenomenon,
  which has no finite dimensional equivalent, is due to the appearance
  of some anomalous dissipation mechanism which transports energy to
  infinity. This prevents the energy from building up locally and
  allows the system to converge to the invariant measure.  The
  invariant measure is constructed explicitly and some of its
  properties are analyzed.
\end{abstract}

\maketitle

\section{Introduction}
\label{sec:intro}

Consider the infinite dimensional linear system of coupled stochastic
differential equations
\begin{equation}
  \label{InviscidEquation}
  \left\{
    \begin{aligned}
      \dot a_n(t) &= a_{n-1}(t) - a_{n+1}(t) +
      \delta_{n,1} \dot W(t)\qquad n\in \NN\\
      a_0(t)&=0,
    \end{aligned}
  \right.
\end{equation}
where the forcing, $\dot W(t)$, is additive white noise and
$\delta_{n,1}$ denotes the Kronecker delta. Formally, the unforced
system appears to conserve energy:
\begin{equation}
  \label{energyconservation}
  \frac{d}{dt}\left( \frac12 \sum_{n\in\NN} a_n^2 \right)
  = \sum_{n\in\NN} a_n \dot{a}_n 
  =\sum_{n\in\NN} a_n (a_{n-1} - a_{n+1}) 
  = \sum_{n\in\NN} a_n a_{n-1} - \sum_{n\in\NN} a_n a_{n+1} =0. 
\end{equation}
Of course, this calculation is only formal as the rearrangement of the
summations is justified only if the sequences involved are absolutely
convergent.  Since the matrix which encodes the coupling
in~\eqref{InviscidEquation} is real antisymmetric, any finite even
dimensional truncation of this matrix has the same real Jordan
canonical form as a system of uncoupled simple harmonic oscillators. 

Our goal in this paper is to address the following natural question:
What is the long term behavior of~\eqref{InviscidEquation}? More
precisely, does the forced system~\eqref{InviscidEquation} have an
invariant measure or statistical steady state? Finite dimensional
truncations might lead one to conjecture a negative answer to this
question. This is due to the fact that, in the finite dimensional
setting, there is no mechanism for energy dissipation. The infinite
dimensional system, on the other hand, does indeed give rise to a
nontrivial invariant probability measure which is not supported in
$\ell_2$. Since there is no apparent dissipative mechanism, we loosely
refer to this phenomena as anomalous dissipation. The present authors
have introduced a formal version of this notion in~\cite{Anomalous}
and have used it to analyze a variety of models.

The invariant probability measure described above is explicitly
computable and is given by the formula:
\begin{align}
  \label{InviscidInvariantMeasure}
  a_{n}(t) = \sqrt{\frac{2}{\pi}}\int_{-\infty}^t G_n^0(t-s) dW(s),
\end{align}
where
\begin{align*}
  G_n^0(s)= \int_{-1}^1 \overline{\u_{n-1}(z)} \sqrt{1-z^2} e^{2 i z 
     s} dz,
\end{align*}
$\u_n(z)=i^n U_n(z)$, and $U_n(z)$ is the $n^{th}$ normalized
Chebychev Polynomial of the second kind~\cite{AbromowitzStegun,
  Szego}.

In this paper we show that~\eqref{InviscidEquation} does have a
nontrivial invariant probability measure given
by~\eqref{InviscidInvariantMeasure}. We derive the explicit
representation for this measure and compute its covariance structure.
We also describe a natural class of initial data whose long term
dynamics converge to~\eqref{InviscidInvariantMeasure}. The analysis is
technically elementary and involves only classical orthogonal
polynomials and basic facts from stochastic calculus. The model
analyzed here should be compared to the related models presented
in~\cite{Anomalous} whose structures are similar except that the
coefficients of the $a_{n-1}$ and $a_{n+1}$ terms depend on $n$. Those
models also exhibit anomalous dissipations, but the qualitative
features of their invariant measures differ from those
of~\eqref{InviscidInvariantMeasure} (in particular, the models
in~\cite{Anomalous} have a different covariance structure
than~\eqref{InviscidInvariantMeasure}). The analysis performed
in~\cite{Anomalous}, which is based on generating functions, is also different from the one for~\eqref{InviscidInvariantMeasure}
presented below.

\section{Invariant Measure}
\label{ssec:damped}

We first analyze a damped version of~\eqref{InviscidEquation} in the
interest of guessing the inviscid invariant
measure~\eqref{InviscidInvariantMeasure}. Next, we prove that the
damped invariant measure converges to the inviscid invariant
measure~\eqref{InviscidInvariantMeasure}. Consider the damped
stochastically forced infinite dimensional system:
\begin{equation}\label{eq:ChebSDE}
  \left\{
    \begin{aligned}
      \dot a_n(t) &= a_{n-1}(t) - a_{n+1}(t)  - \nu  a_n(t) +
      \delta_{n,1} \dot W(t)\\
      a_0(t)&=0,
    \end{aligned}
  \right.
\end{equation}
where $\nu\ge 0$ is a parameter. We show that this equation has a
stationary solution given by 
\begin{align}
  \label{eq:ChebStationary}
  a_{n}^\nu(t) = \sqrt{\frac{2}{\pi}}\int_{-\infty}^t G_n^\nu(t-s) dW(s)
\end{align}
where
\begin{align*}
  G_n^\nu(s)= \int_{-1}^1 \overline{\u_{n-1}(z)} 
  \sqrt{1-z^2} e^{-(\nu - 2 i z) s} dz.
\end{align*}

Before establishing that~(\ref{eq:ChebStationary})
solves~(\ref{eq:ChebSDE}), we show that~\eqref{eq:ChebStationary}
is a well defined random variable. It is sufficient
to prove that $\int_0^\infty \big|G_n^\nu(s)\big|^2 ds < \infty$. This
is clear for the case $\nu>0$. We present a short argument for the
case $\nu=0$. Let $f^\gamma_n(z)= \u_n(z) \sqrt{1-z^2} e^{2i \gamma
  z}$ where $\gamma \in [0,\pi]$. Note that $\int_{-1}^1 |
f_n^\gamma(z)|^2 dz < \infty$ and independent of $\gamma$. Also note
that $G_n^0(\gamma + k \pi) = \int_{-1}^1 f^\gamma_{n-1}(z) e^{2 \pi i
  k z} dz$ is the $-k^{th}$ Fourier coefficient of $f_{n-1}^\gamma$.
The Plancharel
theorem and compactness of $[0,\pi]$ imply that
\begin{align*}
  \sum_{k \in \ZZ} \big| G_n^0(\gamma+k \pi)|^2 < \mathcal{C}<\infty
\end{align*}
for some constant $\mathcal{C}$ which is independent of $\gamma$.
These observations give the desired estimate
\begin{align*}
  \int_0^\infty \big[G_n^0(s)\big]^2 ds &\leq \sum_{k \in \ZZ}
  \int_0^\pi\big|G_n^0(\gamma + k \pi)\big|^2 d\gamma =\int_0^\pi
  \left[ \sum_{k \in \ZZ} \big| G_n^0(\gamma + k
    \pi)\big|^2 \right] \, d\gamma\\
  &= \int_0^\pi \left[ \int_{-1}^1 |f_{n-1}^\gamma(z)|^2 dz
  \right]\,d\gamma = \pi \|f_{n-1}^\gamma\|_2^2 \leq \mathcal{C} \pi <
  \infty
\end{align*}
and show that~\eqref{eq:ChebStationary} is a well
defined random variable. In fact, equation~\eqref{eq:ChebStationary}
defines an infinite dimensional Gaussian Markov process, $\{ a_n^\nu
(t) \}$. The measure that~\eqref{eq:ChebStationary} induces is clearly
invariant under time shifts of the driving Brownian motion. Thus, once
we show that~\eqref{eq:ChebStationary} is a solution of
equation~\eqref{eq:ChebSDE} for all $\nu \geq 0$ the previous
observation implies that~\eqref{eq:ChebStationary} induces an
invariant measure for~\eqref{eq:ChebSDE} for all $\nu\geq 0$. Since
the invariant Gaussian measure we constructed is explicit, we can
calculate its covariance structure: This will be done in
section~\ref{sec:cov}.

To show that~\eqref{eq:ChebStationary} solves~\eqref{eq:ChebSDE} for
all $\nu \geq 0$, we use the following basic fact about Wiener
integrals: For nice $G$, e.g. $G\in C^1(\mathbb{R})\cap
H_1(\mathbb{R}_+)$, $g(t)=\int_{-\infty}^t G(t-s)\, dW(s)$ satisfies
\begin{align}
  \label{derivative1}
  dg(t) = G(0)\,dW(t) + \Big[ \int_{-\infty}^t G'(t-s)\, dW(s)
  \Big]\,dt.
\end{align}
Denote by $\mu(dz)$ the measure $\sqrt{1-z^2}dz$. $G_1^\nu (0) =
\sqrt{\frac{\pi}2}$ and the orthogonality of the $\{ \u_n \}$ in
$L^2([-1,1], \mu)$ imply that $G_n^\nu (0)=0$ for all $n>1$.
Calculating the ${G_n^\nu}^\prime (t-s)$ and using the recurrence
relation~\eqref{recurrence}, we arrive at the relation
\begin{equation}
  \label{derivative2}
  {G_n^\nu}^\prime (t-s) = - \int_{-1}^1 (\nu - 2iz  ) \overline{\u_{n-1}(z)} 
  \sqrt{1-z^2} e^{-(\nu - 2iz)(t-s)} dz 
  = G_{n-1}^\nu(t-s) - G_{n+1}^\nu (t-s) -\nu G_n^\nu (t-s).
\end{equation}
We recover the equations in~\eqref{eq:ChebSDE} by
applying~\eqref{derivative1} and ~\eqref{derivative2} to
formula~\eqref{eq:ChebStationary} for the sequence $\{ a_n^\nu \}$ and
using the recurrence relation
\begin{align}
  \label{recurrence}
  \u_{n+1}(z)-\u_{n-1}(z)&= 2i z \u_n(z)\ .
\end{align} 
Relation~\eqref{recurrence} follows directly from the well-known three
term recurrence relation for the Chebychev polynomials:
\begin{align*}
  U_{n+1}(z)+U_{n-1}(z)=2z U_n.
\end{align*}
This shows that~\eqref{eq:ChebStationary} does indeed provide a
solution and stationary measure for the infinite dimensional coupled
system~\eqref{eq:ChebSDE}. \bigskip

Finally, we explain the origin of
formula~\eqref{eq:ChebStationary} for the $\{ a_n^\nu \}$.  Notice
that the coupling matrix of equation~\eqref{InviscidEquation} is the
Jacobi matrix associated to the three term recurrence of the Chebychev
polynomials of the second kind~\cite{Deift, Szego}.  Therefore,
working with generating functions in the Chebychev polynomials will
diagonalize~\eqref{InviscidEquation}. With this in mind, consider
\begin{align*}
   \alpha^\nu(z,t)= \sum_{n=1}^\infty a_n^\nu(t)\u_{n-1}(z).
\end{align*}
Using   the  recurrence  relation~\eqref{recurrence}   and the   facts
$\u_0(z)=\sqrt{\frac2{\pi}}$  and   $\u_1(z)=2iz   \u_0(z)$, one   can
calculate the time derivative of $\alpha^\nu (z,t)$ as follows:
\begin{align*}
  \dot \alpha^\nu(z,t)
  &= \sum_{n=1}^\infty \dot a_n^\nu (t) \u_{n-1}(z)\\
  &=\sum_{n=1}^\infty \left[ a_{n-1}^\nu(t) 
    - a_{n+1}^\nu(t) - \nu a_n^\nu (t)  + \delta_{n,1} \dot W(t)  
  \right] \u_{n-1} (z) \\
  &= -\nu \alpha^\nu (z,t) + \sum_{n=1}^\infty a_{n-1}^\nu (t) \u_{n-1}(z) 
  - \sum_{n=1}^\infty a_{n+1}^\nu (t) \u_{n-1}(z) + \u_0(z) \dot W (t) \\
  &= -\nu \alpha^\nu (z,t) + a_1^\nu (t) \u_1(z) -2iz a_1^\nu (t) \u_0(z) 
  +2iz \alpha^\nu (z,t) + \u_0(z) \dot W (t) \\
  &= -\nu \alpha^\nu (z,t) +2iz \alpha^\nu (z,t) + \u_0(z) \dot W (t).
\end{align*}
This implies that $\alpha^\nu(z,t)$ satisfies the 
stochastic differential equation
\begin{equation}
  \label{auxilliary}
  \dot \alpha^\nu(z,t)= -(\nu +2iz) \alpha^\nu(z,t)
  +\sqrt{\frac{2}{\pi}}\dot W(t)  
\end{equation}
with $z$ viewed as a parameter varying in $[-1,1]$. The solution to the
initial value problem for~\eqref{auxilliary} and $t \geq s$ is
\begin{align}
  \label{auxilliarysolution}
  \alpha^\nu(z,t)=\chr^\nu(t-s,z)\alpha^\nu(z,s) 
  + \sqrt{\frac{2}{\pi}}\int_s^t
  \chr^\nu(t-r,z) dW(r)
\end{align}
where $\chr^\nu(t,z)=\exp{( (2i z-\nu) t)}$. Letting $s\rightarrow
-\infty$ and assuming ``nice initial conditions'' and $\nu>0$, we
obtain the form of the invariant measure in
formula~\eqref{eq:ChebStationary}. We note that although this is the
way in which the expressions above were derived or ``guessed'', none
of the results depend on this derivation.

\section{Covariance Structure of the Invariant Measure}
\label{sec:cov}

We compute the covariance structure
of~\eqref{InviscidInvariantMeasure}, $c^0(m,n)=\mathbb{E}a_m^0 a_n^0$,
by first computing the covariance structure
of~\eqref{eq:ChebStationary}, $c^\nu(m,n)=\mathbb{E} a_m^\nu a_n^\nu$,
and taking the limit as $\nu\rightarrow 0$. The justification of this
procedure requires two short steps: First, show that there is a
sequence of $\nu_k \rightarrow 0$ such that $a_n^{\nu_k} (0)
\rightarrow a_n^0(0)$ almost surely as $k\rightarrow \infty$; second,
show that $a_n^\nu$ is almost surely a uniformly continuous function of $\nu \in
(0,M)$ for any $M>0$. Therefore, the convergence of $a_n^\nu \rightarrow a_n^0$
as $\nu \rightarrow 0$ holds almost surely. The following estimate and
the Borel-Cantelli lemma complete the first step:
\begin{equation}
  \label{BorelCantelli}
  \mathbb{P}\left( |a_n^{\nu_k}(0) - a_n^0(0)|
    > \frac1{k}    \right) \leq 
  k^2 \int_0^\infty |e^{-\nu_k t} -1|^2 |G_n^0(t)|^2 dt.
\end{equation}
To apply the Borel-Cantelli lemma, simply choose $\nu_k$ so that the
sum over $k$ of the right hand side of~\eqref{BorelCantelli} is
finite. For such a choice of $\nu_k$, $a_n^{\nu_k}(0) \rightarrow
a_n^0(0)$ almost surely. To show that $a_n^\nu(0)$ is almost surely a uniformly 
continuous function of $\nu \in (0,M)$, we appeal to the
Kolmogorov continuity theorem and the following estimate. Fix $0<\eta<
\frac12$, $\rho>\nu>0$, and observe that
\begin{align*}
  \mathbb{E} |a_n^\nu(0) -a_n^\rho (0)|^2 &=
  \int_0^\infty e^{-2\nu t} |1-e^{-(\rho - \nu) t}|^2 |G_n^0(t)|^2 dt\\
  &= \int_0^{|\rho - \nu|^{-\eta}} e^{-2\nu t} |1-e^{-(\rho - \nu)
    t}|^2 |G_n^0(t)|^2 dt + \int_{|\rho - \nu|^{-\eta}}^\infty
  e^{-2\nu t}
  |1-e^{-(\rho - \nu) t}|^2 |G_n^0(t)|^2 dt \nonumber \\
  &\leq C_1 | \rho - \nu |^{2(1-\eta)} + C_2 e^{-2\nu |\rho - \nu
    |^{-\eta }} \leq C_3 | \rho - \nu |^{2(1-\eta)},
\end{align*}
where the last inequality holds for $|\rho - \nu|$ sufficiently small
and $C_1,C_2$, and $C_3$ are constants which do not depend on $\eta,
\nu$, and $\rho$. This completes the justification for computing
$c^0(m,n)$ by taking the limit of $c^\nu (m,n)$ as $\nu \rightarrow
0$. \bigskip

We now compute $c^\nu(n,n)$:
\begin{equation}
  \label{variance}
  \begin{aligned}
    \mathbb{E} a_n^\nu(0) \overline{a_n^\nu(0)} &=
    \frac2{\pi} \int_0^\infty  G_n^\nu(s) \overline{G_n^\nu(s)} ds\\
    &= \frac2{\pi} \int_0^\infty \left[ \int_{-1}^1 \overline{\u_{n-1}(z)}
      \sqrt{1-z^2} e^{-(\nu - 2iz)s} dz \right] \overline{\left[
        \int_{-1}^1 \overline{\u_{n-1}(z')}
        \sqrt{1-{z'}^2} e^{-(\nu - 2iz')s}  dz'  \right]} ds \\
    &= \frac2{\pi} \int_{-1}^1 \int_{-1}^1 \overline{\u_{n-1}(z)}
    \u_{n-1}(z') \sqrt{1-z^2} \sqrt{1-{z'}^2}
    \left[ \int_0^\infty e^{-2(\nu - i(z-z'))s}ds \right]dz dz'\\
    &= \frac2{\pi} \int_{-1}^1 \int_{-1}^1 U_{n-1}(z) U_{n-1}(z')
    \frac{\sqrt{1-z^2} \sqrt{1-{z'}^2}}{2\nu -2i (z-z')} dz dz'.
  \end{aligned}
\end{equation}
The fact that the dynamics of $\{ a_n \}$ is real implies that one
needs only to compute the real part of equation~\eqref{variance}:
\begin{equation}
  \frac{\nu}{\pi} \int_{-1}^1 \int_{-1}^1 U_{n-1}(z) U_{n-1}(z') 
  \frac{\sqrt{1-z^2} \sqrt{1-{z'}^2}}{\nu^2 +(z-z')^2} dz dz'.
\end{equation}
Introducing the change of variables $z=\cos(\pi \theta), z'= \cos(\pi
\theta')$ and using the fact that $U_n(\cos(\pi \theta))=
\sqrt{\frac2{\pi}} \frac{\sin(\pi (n+1)\theta)}{\sin(\pi \theta)}$, see
for example~\cite{Szego}, leads to the following integral:
\begin{equation}
  c^\nu(n,n)=2\nu \int_0^1 \int_0^1 \frac{\sin(\pi n z) \sin(\pi n z') 
    \sin(\pi z) \sin(\pi z')}{\nu^2 + (\cos(\pi z) - \cos(\pi z'))^2} dz dz' .
\end{equation}
Analyzing the limit $\nu \rightarrow 0$, one finds that:
\begin{equation}
  c^0(n,n)=\lim_{\nu \rightarrow 0} c^\nu(n,n)
  =  \int_0^2 \sin \left( \pi \frac{\xi}{2} \right) 
  \sin^2\left( \pi \frac{n\xi}{2} \right) d\xi.
\end{equation}
Note that $\lim_{n\rightarrow \infty} c^0(n,n) = \frac2{\pi}$ which
immediately shows that the invariant measure is not supported on
$\ell_2$: We will discuss the implication of this fact  in
Section~\ref{sec:basin}. For the moment, we only remark that the existence of this limit is consistent with the systems' invariance by translation except for the forcing and the boundary condition at $n=0$.

We compute the general covariance structure $c^\nu(m,n)$ in two
steps: $n-m$ is odd; $n-m$ is even. First note that
\begin{equation}
  \label{covariance}
  \mathbb{E}a_n^\nu(0) \overline{a_m^\nu(0)} 
  =  \frac{(-1)^n}{\pi} i^{n+m} \int_{-1}^1 
  \int_{-1}^1 U_{n-1}(z) U_{m-1}(z') \sqrt{1-z^2} \sqrt{1-(z')^2} 
  \frac{\nu + i (z-z')}{\nu^2 + (z-z')^2} dz dz'.
\end{equation}
The cases when  $n-m$ is odd or even must be treated separately:
\bigskip

\noindent {\it Case 1}: If $n-m$ is odd, then
\begin{align*}
  \mathbb{E}a_n^\nu(0) \overline{a_m^\nu(0)} &=   
  \frac{(-1)^n}{\pi} i^{n+m+1} \int_{-1}^1 \int_{-1}^1 U_{n-1}(z) U_{m-1}(z') 
  \sqrt{1-z^2} \sqrt{1-(z')^2} \frac{ (z-z')}{\nu^2 + (z-z')^2} dz dz' \\
  \mathbb{E}a_m^\nu(0) \overline{a_n^\nu(0)} &= \frac{(-1)^m}{\pi}
  i^{n+m+1} \int_{-1}^1 \int_{-1}^1 U_{n-1}(z) U_{m-1}(z')
  \sqrt{1-z^2} \sqrt{1-(z')^2} \frac{ (z-z')}{\nu^2 + (z-z')^2} dz
  dz'.
\end{align*}
Since $n-m$ is odd, these two expressions have opposite signs. On the
other hand, the fact that the dynamics of the $\{ a_n \}$ is real implies that the two expressions must be equal. Therefore, $c^\nu (m,n)=0$  which further implies that $c^0 (m,n) = 0$.
\bigskip

\noindent {\it Case 2}:  If $n-m$ is even, then 
\begin{equation} \label{even}
  \mathbb{E}a_n^\nu(0) \overline{a_m^\nu(0)} 
  =   \frac{(-1)^n}{\pi} i^{n+m} \nu \int_{-1}^1 
  \int_{-1}^1 U_{n-1}(z) U_{m-1}(z') \sqrt{1-z^2} 
  \sqrt{1-(z')^2} \frac{1}{\nu^2 + (z-z')^2} dz dz'.
\end{equation}
Once again, analyzing the limit of~\eqref{even} as $\nu \rightarrow 0$ leads to the formula for $c^0(m,n)$:
\begin{eqnarray}
c^0(m,n)=\lim_{\nu \rightarrow 0} c^\nu(m,n)
  &=& (-1)^n i^{m+n} \int_0^2 \sin \left( \pi \frac{\xi}{2} \right) 
  \sin \left( \pi \frac{m\xi}{2} \right) \sin \left( \pi \frac{n\xi}{2} \right) d\xi \nonumber \\
  &=& (-1)^n i^{n+m} \frac2{\pi} \left[   \frac1{(n+m)^2 -1} - \frac1{(n-m)^2 -1}         \right]
\end{eqnarray}

\section{Basin of Attraction of the Invariant Measure}
\label{sec:basin}

Next, we prove that if the initial condition, $\{ a_n \}$, is in
$\ell_2$, i.e. $\sum_{n=1}^\infty |a_n|^2 < \infty$, then the dynamics
converges weakly to the invariant measure~\eqref{eq:ChebStationary}
for any $\nu\ge 0$: Any finite collection of coordinates,
$a^\nu_{i_1},...,a^\nu_{i_k}$, converges to~\eqref{eq:ChebStationary}
as the initial condition is pulled back to $s= -\infty$. Assume that
$\{ a_n \} \in \ell_2$ and construct the function
\begin{equation}
  \label{initcond}
  \alpha_0^\nu(z)= \sum_{n=1}^\infty a_n \u_{n-1}(z).
\end{equation}
If $\alpha^\nu (z,s) =\alpha_0^\nu(z)$ is the initial condition at
time $s$ for equation~\eqref{auxilliary}, then the solution at time
$t>s$ is given by equation~\eqref{auxilliarysolution}. Note that the
dynamics is well defined for $\ell_2$ initial data and the solution
remains in $\ell_2$ for all $t<\infty$. 
To recover the solution 
at time $t$ we simply use orthogonality of the $\{ \u_n \}$:
\begin{equation}
  \label{solution}
  \begin{aligned}
    a^\nu_n(t)&=\int_{-1}^1 \overline{\u_{n-1} (z)}\alpha^\nu(z,t)
    \sqrt{1-z^2} dz  \\
    &=\int_{-1}^1 \overline{\u_{n-1} (z)} \chr^\nu(t-s,z)\alpha^\nu(z,s)
    \sqrt{1- z^2}dz \\
    & \quad + \int_{-1}^1 \overline{\u_{n-1}(z)} \left[
      \sqrt{\frac{2}{\pi}}\int_s^t \chr^\nu(t-r,z) dW(r) \right]
    \sqrt{1-z^2} dz.
  \end{aligned}
\end{equation}
Since $\overline{\u_{n-1} (z)} \alpha_0^\nu(z) \sqrt{1- z^2}
\chi_{[-1,1]}(z)\in L^1(\mathbb{R})$, standard Fourier analysis
implies that the first integral vanishes as $s\rightarrow -\infty$.
The second integral converges to the form of the invariant
measure~\eqref{eq:ChebStationary}. The convergence is uniform if a
finite collection of coordinates is fixed. Therefore, for any bounded
cylinder function, convergence is established, which, in turn,
establishes weak convergence of solutions with $\ell_2$ initial data
to the invariant measure~\eqref{eq:ChebStationary}. 

A drawback of the convergence result above is that, when $\nu=0$, the
invariant measure is not supported on $\ell_2$. Hence, we may wonder
about convergence of initial data which is not in $\ell_2$ and is in
the support of the invariant measure or even in $\ell_\infty$. This
question, however, turns out to be quite complicated as the behavior
of the initial value problem for~(\ref{InviscidEquation}) depends
sensitively on the initial condition. Since ~(\ref{InviscidEquation})
is linear, it suffices to understand the solution of the unforced
system with initial condition $a_n(s) = a_n^0$:
\begin{equation}
  \label{IVP}
  \left\{
    \begin{aligned}
      \dot a_n(t) &= a_{n-1}(t) - a_{n+1}(t), \qquad a_n(0)= a^0_n
      \qquad n\in \NN\\
      a_0(t)&=0.
    \end{aligned}
  \right.
\end{equation} 
The solution of the forced system with initial condition $a_n(s) =
a_n^0$ is then obtained by adding the solution
of~(\ref{InviscidEquation}) with $a_n(s)=0$ to that of~(\ref{IVP}).
Cataloging the behavior of the solutions of~(\ref{IVP}) is
complicated, even if we restrict ourselves to initial condition
in~$\ell_\infty$. The difficulty is immediately understandable if one
notices that
\begin{equation}
  \label{eq:fixedpoint}
  a_{2n} = 1, \qquad a_{2n+1} = 0, \qquad n \in \NN 
\end{equation}
is a fixed point for~(\ref{IVP}) and belongs to~$\ell_\infty$.
Similarly, one can find time-periodic solutions with arbitrary period
$T>0$ which also belong to~$\ell_\infty$. For the sake of brevity, we
will refrain from attempting a complete analysis of~(\ref{IVP}). We
note, however, that such an analysis has been performed in detail
in~\cite{Anomalous} for models related to~(\ref{IVP}).  \bigskip

\noindent {\bf \it Acknowledgments}:  
We thank Percy Deift, Charles Fefferman, Stephanos Venakides, and Xin
Zhou for useful conservations. J. Mattingly is supported in part by
the Sloan Foundation and by an NSF CAREER award DMS04-49910. T. Suidan
is supported in part by NSF grant DMS05-53403. E. Vanden-Eijnden is
supported in part by NSF grants DMS02-09959 and DMS02-39625, and by
ONR grant N00014-04-1-0565.


\end{document}